# Plasma focus based repetitive source of fusion neutrons and hard x-rays

V Raspa*[1], F Di Lorenzo[1], P Knoblauch[1], A Lazarte[1], A Tartaglione[2], A Clausse[3] and C Moreno[1]

Address: [1]Laboratorio Plasma Focus – Instituto de Física del Plasma – Departamento de Física, FCEyN – Universidad de Buenos Aires – PLADEMA, Buenos Aires, Argentina, [2]IB – CAB – CNEA, Bariloche, Argentina and [3]PLADEMA – UNICEN – CNEA, Tandil, Argentina

Email: V Raspa* - raspa@df.uba.ar; F Di Lorenzo - fjdl@yahoo.com; P Knoblauch - pablotk@df.uba.ar;
A Lazarte - alelazarte@yahoo.es; A Tartaglione - tartagla@cab.cnea.gov.ar; A Clausse - clausse@exa.unicen.edu.ar;
C Moreno - moreno@df.uba.ar

* Corresponding author

## Abstract

A plasma focus device capable of operating at 0.2 pulses per second during several minutes is used as a source of hard x-rays and fast neutrons. An experimental demonstration of the use of the neutrons emissions for radiation probing of hydrogenated substances is presented, showing a particular application in detecting water concentrations differences in the proximity of the device by elastic scattering. Moreover, the device produces ultrashort hard x-rays pulses useful for introspective images of small objects, static or in fast motion, suitable for the identification of internal submillimetric defects. Clear images of metallic objects shielded by several millimeters iron walls are shown.

## Background

Pulsed sources of fusion neutrons and x-rays, as well as energetic electron and ion beams can be produced by means of plasma focus devices. After its invention in the 60's [1,2] they were intensively studied as nuclear fusion devices and were described in detail, among other references, in [3-5]. During the last decade, they began to be investigated as convenient sources for several technological applications, to be cited below, of the mentioned radiations.

Essentially these devices are plasma accelerator guns, where the plasma is generated in a low pressure (1–10 mbar) atmosphere by a pulsed powerful capacitive discharge between a pair of coaxial cylindrical electrodes. The Lorentz force drives a plasma sheath along the electrodes, which radially collapses at the symmetry axis forming a hot dense plasma pinch of about 100 ns.

Intense pulses of electromagnetic radiation as well as ions and electrons beams are emitted as a result of the focusing process. The electromagnetic emission from the collapsing plasma has a very high brightness and exhibits a broadband spectra ranging from visible light to soft x-rays. Fusion reactions can be obtained if deuterium or deuterium-tritium mixtures are used, with the consequent emission of fast neutrons. The kinetic energy of the emitted neutrons being 2.45 MeV in the first case and 2.45 MeV or 14.1 MeV in the second one. Both emissions are almost monochromatic [3-5].

Since the neutron and also the hard x-ray burst duration of this source is in the 10–100 ns range, and that its emission can be turned off, the plasma focus becomes an interesting alternative to commercially available radioisotopic sources of both neutrons and hard x-rays.

Due to the high effective cross section of hydrogen for neutron dispersion, plasma focus neutron pulses can be used as radiation probes to detect hydrogenated substances by means of neutron scattering. Examples of potential applications of this radiation are soil humidity studies [6] and detection of hidden dangerous or illegal substances including drugs, weapons and plastic explosives [7]. The plasma focus neutron emission is also suitable for dark matter research [8], radioisotopes production [9] and neutron therapy [10].

Other reported applications are radiographies of biological samples [11-13], small pieces made of different materials [14], and a rotating car wheel [15]. Introspective images of small metallic pieces [16,17] even through several millimeter thick metallic walls [18] were also produced. Additionally, a radiography of an aluminum turbine rotating at 6120 rpm using a plasma focus pulse was reported [19]. The tomographic reconstruction of both surface and volume of small metallic objects was investigated as another non-conventional imaging application of the hard x-ray emission of plasma focus devices [16,20]. The spatial resolution of the digitalized images has demonstrated to be suitable for 3-dimensional tomographic reconstruction of an object with just 8 projections.

Other capabilities of the plasma focus emissions include lithography with ten nanometers resolution [21,22], medical radiation therapy applications [23], industrial non-destructive testing [24,25], as well as, thin film deposition [26].

Compact and portable plasma focus are being developed to work efficiently on the field at relative low costs when compared to nuclear reactors or linear accelerator based sources. Moreover, the feasibility of combining neutron and x-ray scannings simultaneously in a single device is a unique advantage of plasma foci.

However, plasma foci still presents several challenges that must be overcome in order to extend the uses of these devices as x-ray and neutron source for commercial and industrial applications. In order to be useful over the widest range of applications, a plasma focus device should be able to operate at a discrete repetition rate [27] in order to produce average neutron emission rates about ($10^7$ – $10^{10}$) neutrons/sec, and intense penetrating x-rays beams. Lee *et al* presented a 16 Hz plasma focus operating in neon [21]. More recently, Rapezzi *et al* developed a mobile repetitive device for industrial applications [28].

In the current paper a repetitive plasma focus device operating with deuterium aimed to a pulsed source of neutrons and hard x-rays is presented. The repetition rate is controlled by means of a variable clock synchronized with a triggering system and a power supply, which ensures a regular operation up to 0.2 Hz. Moreover, the feasibility of technological application are analyzed.

**Device design and output characteristics**

A 4.7 kJ small chamber Mather-type plasma focus was used as repetitive radiation source. The gas chamber was filled with 4.0 mbar of an admixture of 2.5% (in volume) of argon in deuterium. The cylindrical chamber, 1 $dm^3$, is made of a 3-mm thick stainless steel tube. The 2-mm thick front of the chamber is a stainless steel disk, 100 mm OD, used as the hard x-ray emission window. The electrodes are concentric cylinders, 85 mm length, 38 and 73 mm OD respectively, made of hollow electrolytic copper and twelve 3-mm diameter brass rods, respectively. The anode base is made of brass. A 50-mm OD cylindrical Pyrex insulator, 4 mm thick, 30 mm length, is located covering the anode at the base. The chamber design is optimized for hard x-ray and fast neutron production [29]. The footprint of the whole device is 0.60 $m^2$, its height being 1.3 m.

The capacitor bank is composed by 3 modules of six 0.7 $\mu$F Maxwell capacitors. The bank was charged using a 10 kW Maxwell CCDS power supply. The modules were connected in parallel to the discharge chamber through 3 Maxwell spark gaps (model 40264), which were triggered simultaneously by means of a car ignition coil.

The device was able to operate regularly at 30 kV at a repetition rate up to 0.2 Hz during runs of 2 minutes maintaining the temperature within reasonable ranges. The external temperature of the anode reached 30°C above room temperature after each run. The continuous compressed air flow needed to set the spark-gap operating voltage and cleaning, was sufficient to cool this element. The external surface of the spark-gap plastic encasing body heated up only few degrees over the room temperature, whereas the chamber refrigerated by natural air convection with the environment.

A microprocessor-based control system synchronized both the power supply and the triggering system (Fig. 1). The peak current delivered was in every shot ~360 kA. Higher frequencies are in principle possible to achieve introducing a cooling system of the chamber and the electrodes. The higher frequency achieved during the tests was 1 Hz during 20 seconds.

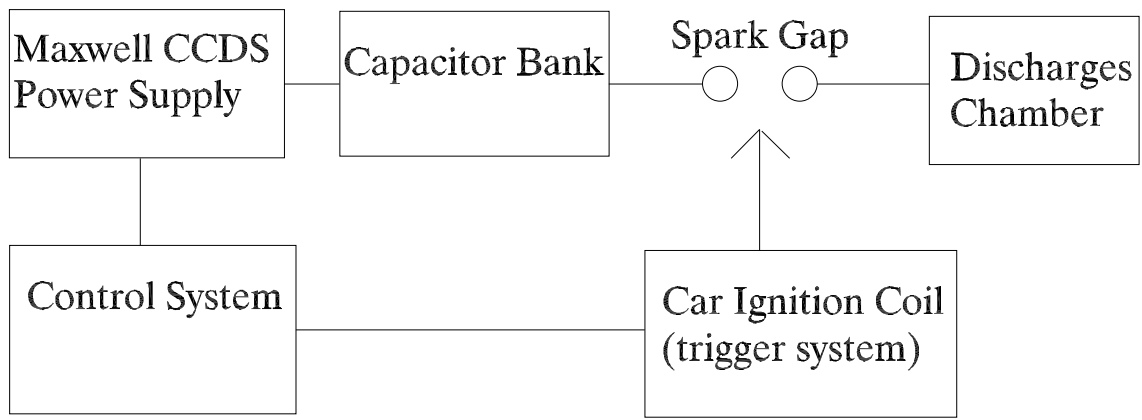

**Figure 1**
Operation scheme of the synchronized charging and triggering systems.

The discharge current was monitored by a non-integrating Rogowski coil. The x-ray emission was detected with a NE102A scintillator optically coupled to a photomultiplier (PMT) polarized at 800 V and placed 3.9 m away from the chamber. The Rogowski coil and the PMT signals were acquired with a Tektronix TDS540A oscilloscope. Both the PMT and the oscilloscope are placed inside a Faraday cage.

Figure 2 shows the Rogowski coil and PMT signals acquired during a 24 shot sequence at 0.2 Hz. It can be observed that both the amplitude and the first quarter of period of the Rogowski coil signal, are repetitive from shot to shot within ± 50 kA/$\mu$s and ± 50 ns, respectively. The pinch time occurs at (1.48 ± 0.12) $\mu$s, close to the peak current time, which is (1.27 ± 0.05) $\mu$s. After integrating the Rogowski coil signal, it results that the current at pinch is always above 90% of the maximum current value.

As it usually happen in plasma focus discharges focusing not always occur in every shot, specially when working in rep-rate mode without replacing the working gas. In the case shown in figure 2, 4 shots out of 24 failed to focus, whereas very intense current dips were attained in 16 of the others. The remaining 4 focalizations were not particularly intense. The PMT signals show that x-ray pulses are produced around 35 ns after the pinch, lasting each about 50 ns FWHM.

The neutron yield was determined by silver activation [30]. The detector consists of a 30 × 30 × 15 cm$^3$ rectangular box filled with paraffin (see figure 3), in which four Victoreen 1B85 Geigers, each surrounded by a 300 $\mu$m thick silver foil and polarized to 800 V, are allocated. The Geigers are placed, respectively, inside four holes (2 cm in diameter by 9 cm in depth) drilled on the diag-onals of one of the 30 × 30 cm$^2$ faces, at 9 cm from the face center. The amount of paraffin allows for the necessary neutron thermalization to match the silver resonance peak for neutron capture, with a beta decay, located at 5 eV. The background counts of this detector are 66 ± 8 in a time interval of 20 sec.

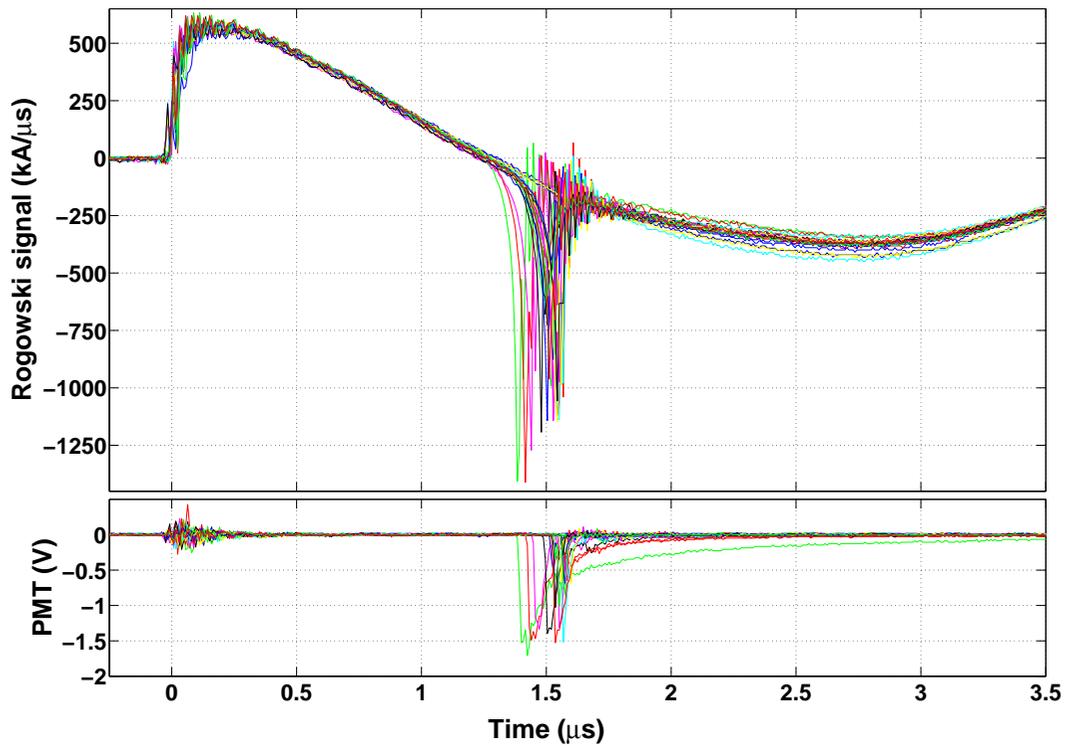

**Figure 2**
Rogowski coil and PMT signals acquired during a 24 shot sequence.

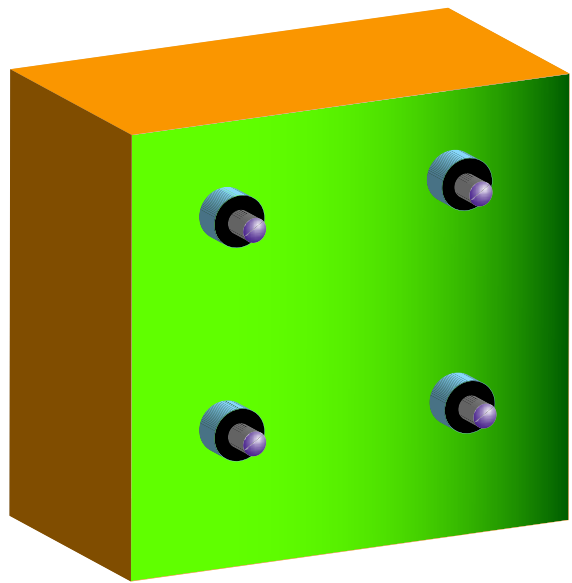

**Figure 3**
Scheme of the fast neutron detector. It is a 30 × 30 × 15 cm$^3$ paraffin block with four Victoreen 1B85 Geigers, each one surrounded by a 300 μm thick silver foil.

Figure 4 shows the measured neutron yield per shot recorded during four independent runs lasting 3 to 5 minutes of 0.2 Hz operation. The average neutron yield is about $10^8$ neutrons per shot during the first 20 shots. After that the yield gradually decreases until it reaches half that number around the shot 60. This may be in part due to the build up of impurities in the filling gas along the run, which slows down the current sheath, resulting in a pinch time delay and a corresponding decrease of the pinch current, or, due to a degradation of the focusing efficiency caused by such impurities. Each run was done with the same gas filling, but it was fully refreshed before starting the run.

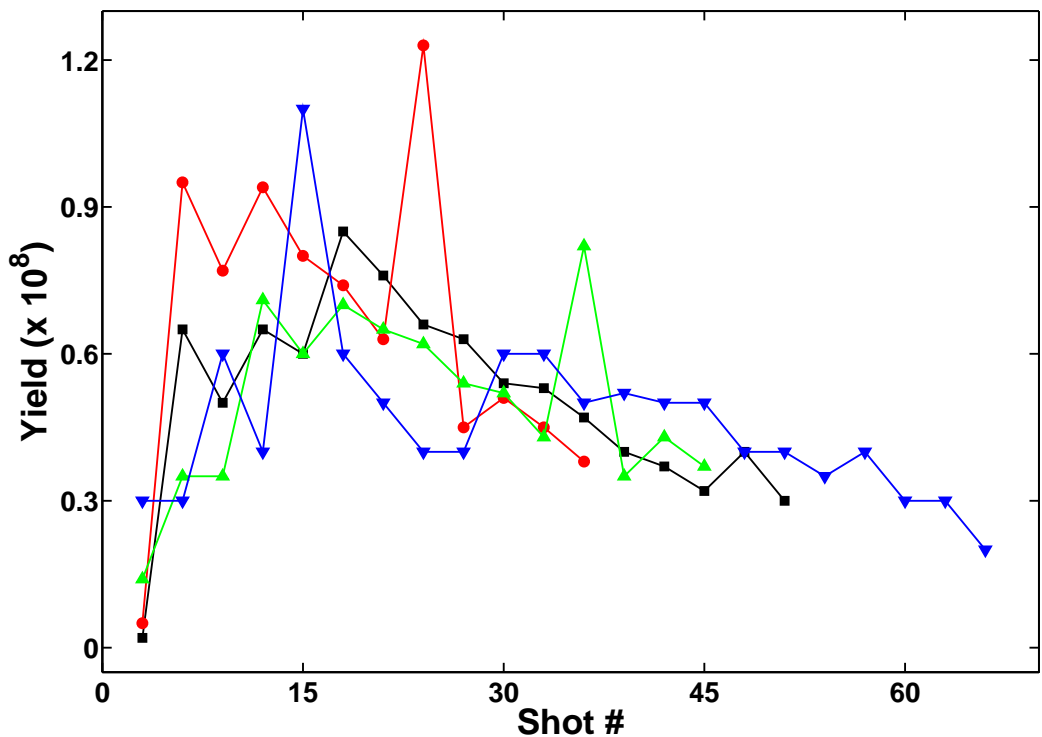

**Figure 4**
Neutron yield for each shot during several runs between 3 and 5 minutes long.

## Applications: Methods and Results

Both neutron and x-ray emissions of the plasma focus device were tested for different technological applications with a potential industrial or therapeutical use. The following sections describe the obtained results.

### *Prospection by fast neutrons scattering*

The neutrons pulses emitted by plasma focus can be used to determine the presence and after the proper calibration estimate the concentration of hydrogenated substances in the proximities of the device [7,31]. The basic idea of the method is depicted in figure 5. A silver activation counter like the one described in section 1, is used as a reference detector to register the neutron yield in each pulse. Part of the emitted neutrons are scattered by the target, if present, and registered by a second detector. This second detector does not contain paraffin for it to be sensitive to scattered slow neutrons and, at the same time, to be insensitive to the direct plasma focus neutron radiation. Moreover, it is shielded with Borax on every side but one to filter slow neutrons not coming through its window. Figure 6 shows a diagram of the detector. It is important to note that this is a side-on diagnostic, that is, the source and the detection system are on the same side respect to the substance to probe.

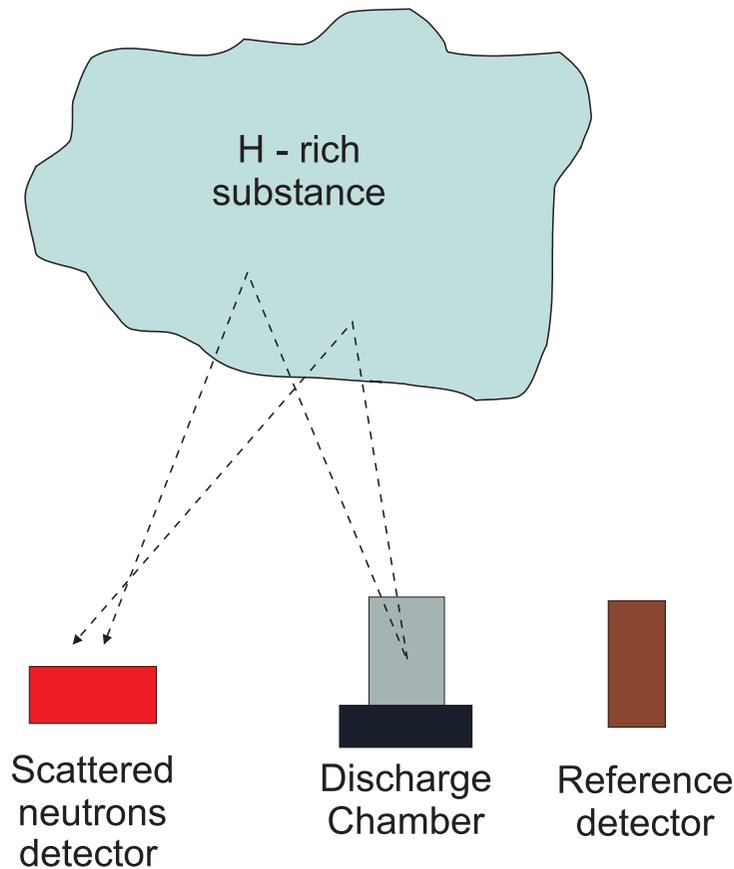

**Figure 5**
Conceptual scheme for the detection of hydrogenated substances by neutron scattering. The scattered neutrons are monitored by a directional detector (scattered neutrons detector). The reference detector, which is like that shown in figure 3, is used to monitor the neutron yield per shot.

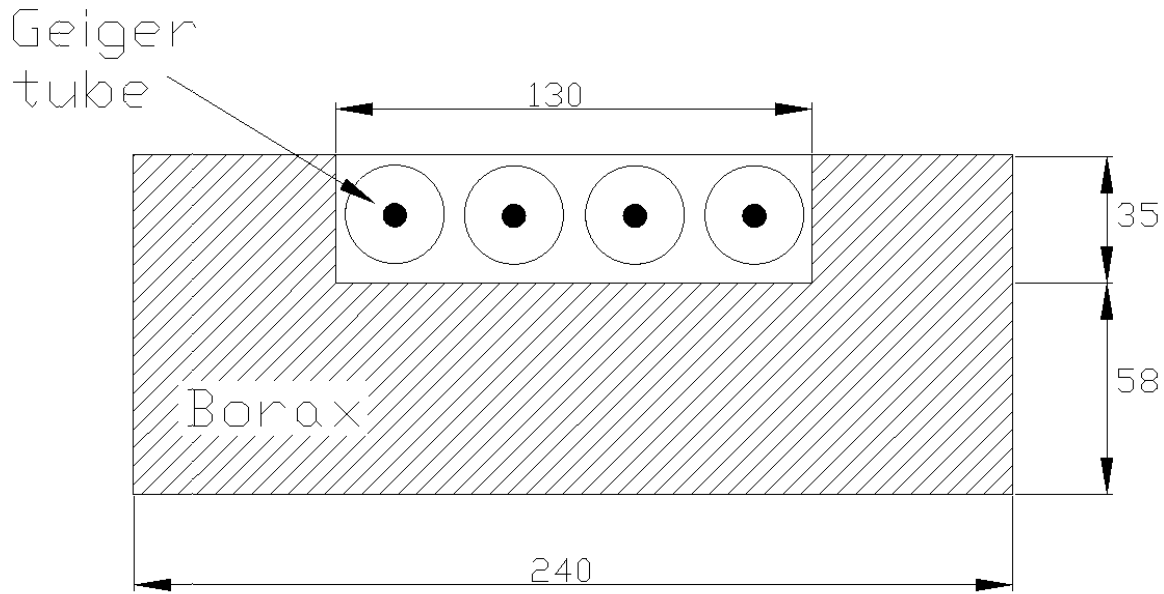

**Figure 6**
Drawing of the scattered neutrons detector. Dimensions are in millimeters.

The feasibility of this kind of application using the present device was tested interrogating a wooden rack with 56 plastic bottles (500 $cm^3$ each) placed close to the plasma focus chamber. Four cases were analysed: a) all the bottles empty, b) 16 filled with water c) 32 filled and d) all of them filled (see figure 7).

Figure 8 shows plots of slow neutron against fast neutron counts for different configurations with their respective linear fits. The slopes, $k$, allow to discriminate between different amounts of water located at few meters of the plasma focus chamber. The experimental $k$ values, $k_e$, resulted in: (0.161 ± 0.011), (0.207 ± 0.011), (0.2838 ± 0.0092) and (0.348 ± 0.013) for cases a) to d), respectively.

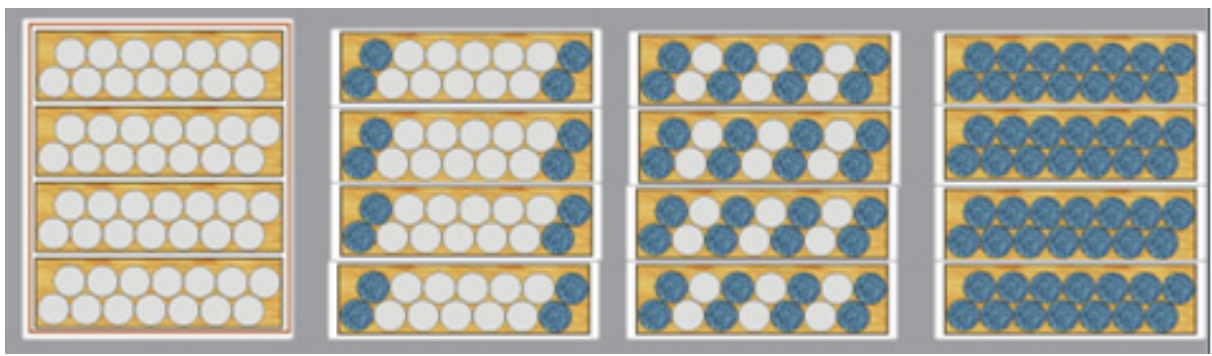

**Figure 7**
The four different configurations of empty and full filled bottles considered to test the system for water amounts detection. From left to right, there are: 0, 16, 32 and 56 water filled bottles, respectively.

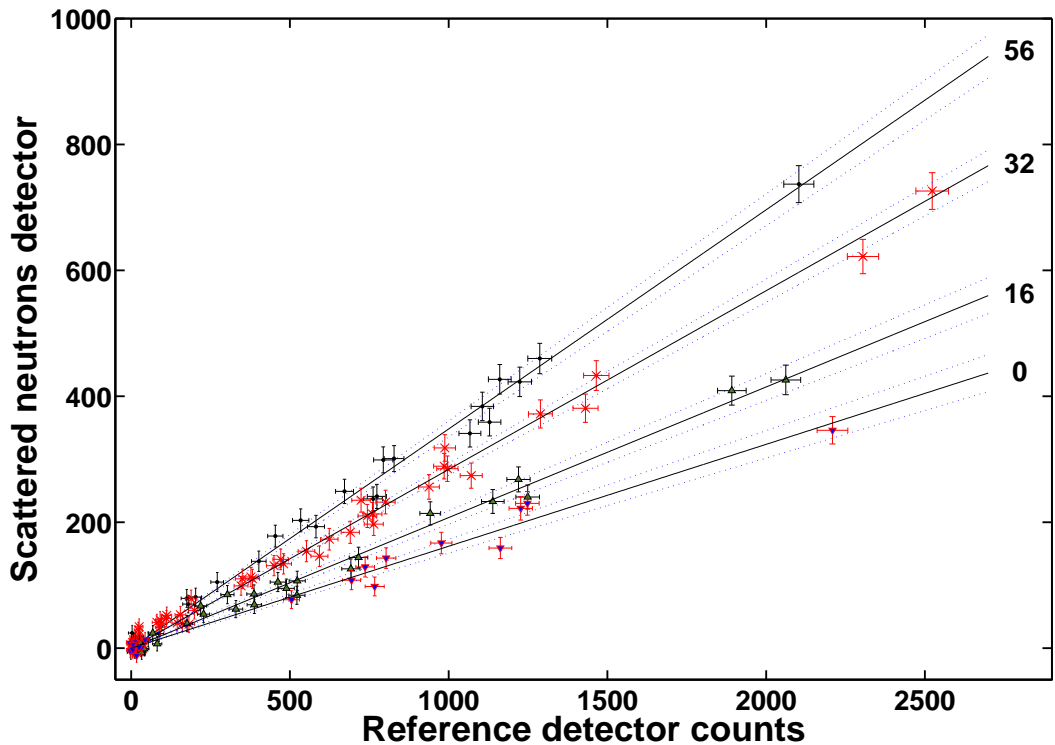

**Figure 8**
Scattered and direct neutron counts for the different configurations of bottles shown in figure 7. The number on the right of each linear fit refers to the number of filled bottles.

The experimental values were checked against a MCNP simulation of the experiment. The simulation included: the bottles, the laboratory floor, ceiling and walls, the capacitor bank, the wooden rack and both detectors. Figure 9 shows a scatter plot of the experimental values against those numerically simulated $k_n$. An acceptable agreement between both quantities is found. The dot-dashed line included in the plot represents complete agreement: $k_e = k_n$. The error bars in $k_n$ correspond to uncertainties in the materials composition of the laboratory building.

### *Hard x-rays introspective imaging of metallic pieces*
Plasma focus operated in single shot mode has demonstrated to be a hard x-ray source suitable to obtain introspective images of small metallic objects.

Commercial AGFA Curix Ortho Regular cassette – film system was used together with standard developer and fixer recommended for that film. No special procedures other than those indi-cated by the manufacturer were required to process the films. Figure 10 illustrates the imaging setup. The objects and the film are placed, on the electrode axis, outside the stainless steel chamber, about 1 m away from the chamber front wall.

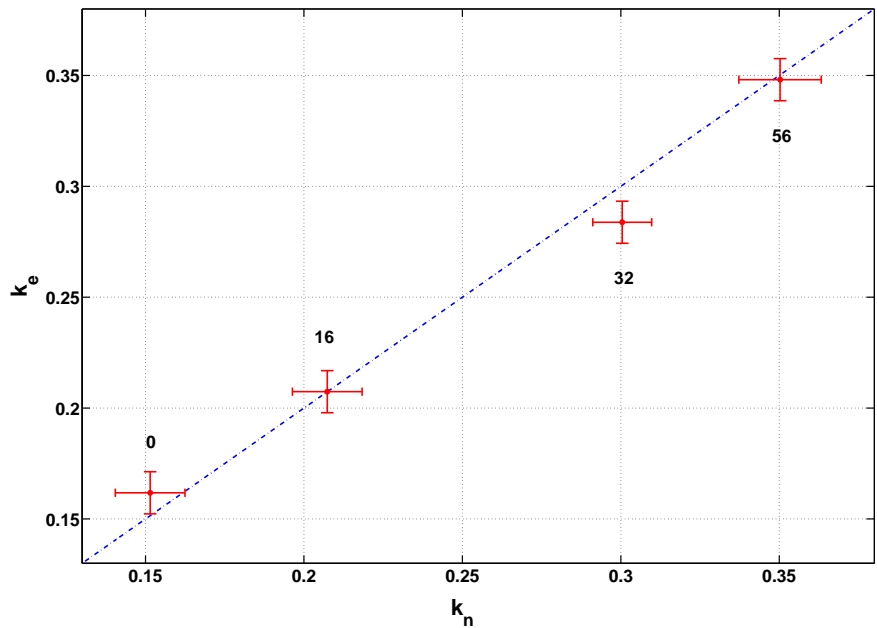

**Figure 9**
Scatter plot of the experimental slopes $k_e$ against their corresponding numerical simulations $k_n$. The dot-dashed line correspond to $k_e = k_n$.

Figure 11(a) shows a digitized radiograph of a metallic BNC tee connector imaged with a single shot through a 9 mm thick iron flange. The dark circle near the upper left of the block is a through hole made on the flange, close to its border. For comparison, Fig. 11(b) shows the same piece radiographied without interposing the flange.

To determine the spatial resolution of the radiographs a 10 mm thick aluminum block screwed with two 1/4–20 bolts (made of brass and steel, respectively) was radiographied. Figure 12 shows the radiograph were submillimetric details of the threads can be easily distinguished (1.27 mm per turn). The overall spatial resolution is 1/6 mm (6 pixels of the scanned image per millimeter measured on the object).

The degree of detail shown in figures 8 and 12 implies that the source size is well below the millimetric scale, since otherwise the images would not show sharp borders with such good contrast.

Finally, in figure 13 snapshots of a strongbox containing the BNC tee and a brass bolt are shown. A single shot radiograph of the set is shown in figure 14.

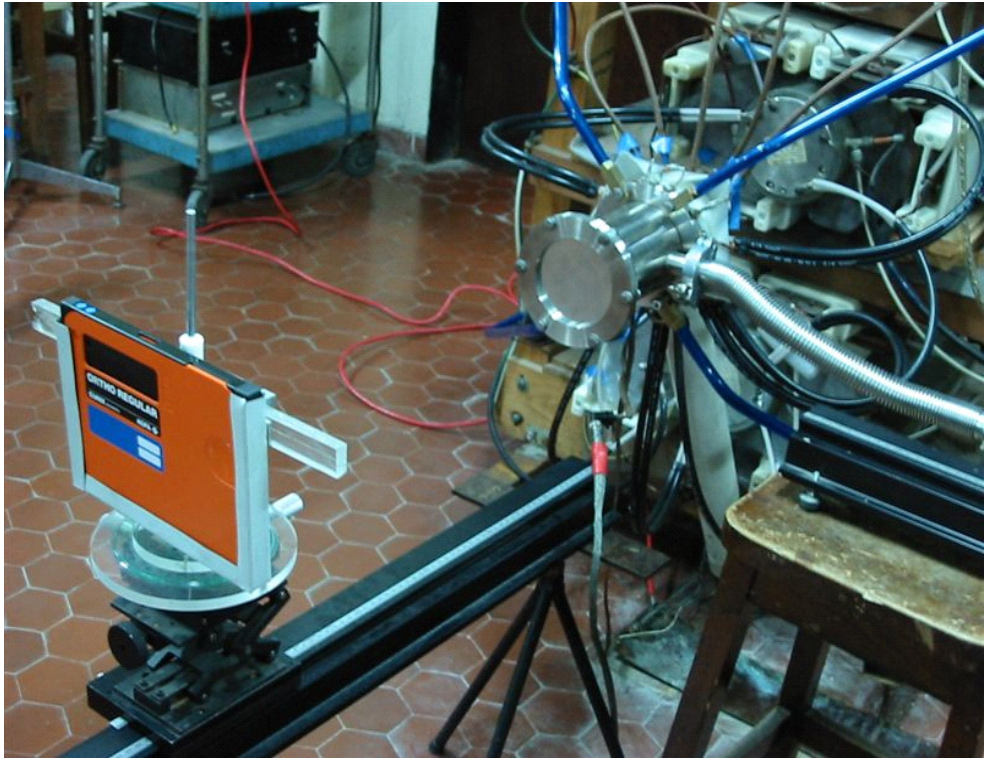

**Figure 10**
Imaging setup. The film is placed outside the discharge chamber, on the electrodes axis, about 1 m away from the chamber front wall.

## Discussion

A repetitive plasma focus device capable of emitting fusion neutrons and hard x-rays was presented, showing that it can be operated at a moderate repetition rate for several minutes. The performance and thermal conditions of the device is stable during 2 to 3 minute runs. Higher shot frequencies can in principle be achieved if some thermal management is provided to cool the heat dissipated in every shot.

It was demonstrated that the emitted neutrons can be used to detect the presence of water near the discharge chamber by neutron scattering. In principle, this procedure can be extended to detect other hydrogenated substances such as explosives or drugs.

Radiographic images could be obtained from the hard x-ray emissions showing submillimetric spatial resolutions with expositions times around 50 ns. The energy and intensity of the x-rays are sufficiently high for the inspection of metallic objects located behind or inside several millimeters of iron or steel. These characteristics are suitable to develop non-intrusive detection systems of internal defects and the imaging of fast rotating components.

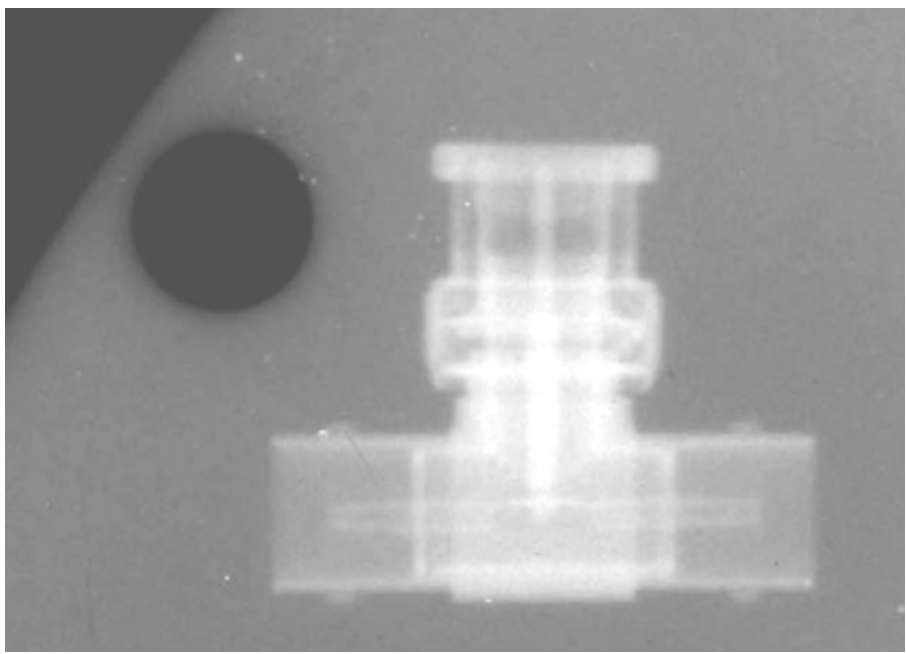

(a)

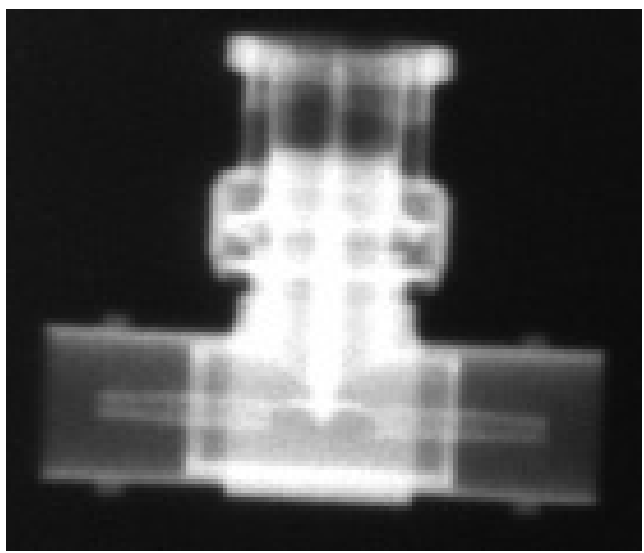

(b)

**Figure 11**
(a) Digitized radiograph of a metallic BNC tee connector imaged with a single plasma focus shot, through a 9 mm thick iron flange. The dark circle near the upper left of the block is a through hole made on the flange, close to its border. (b) The same connector imaged without interposing the flange.

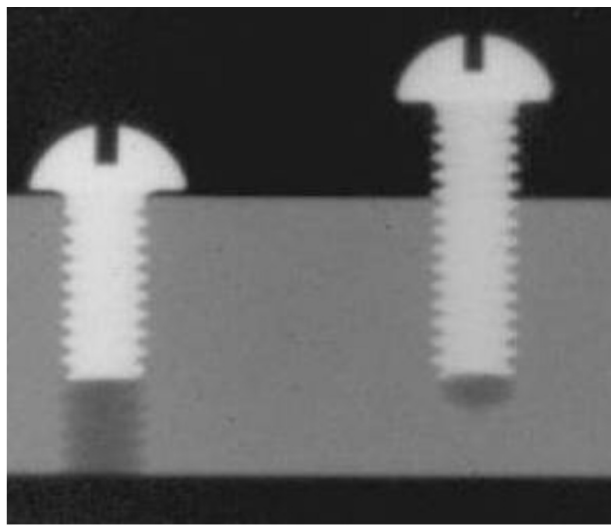

**Figure 12**
A 10 mm thick aluminum block with two 1/4–20 bolts, one made of steel (left) and the other of brass (right).

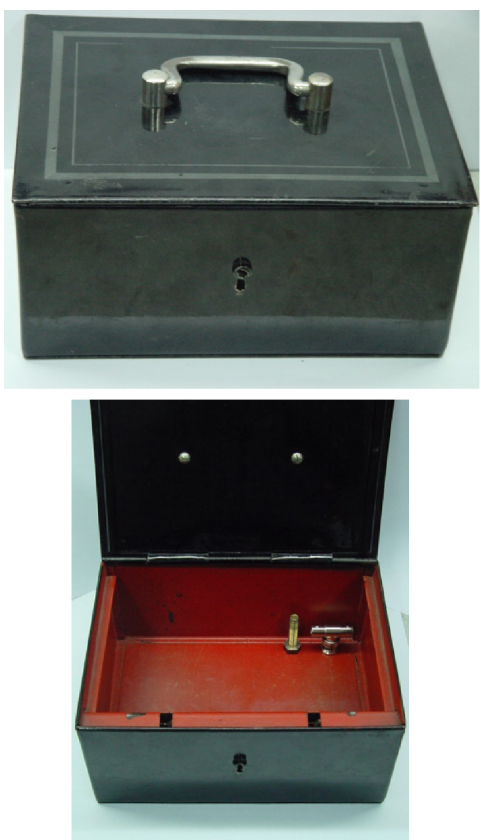

**Figure 13**
Snapshot of a strongbox containing a BNC tee and a brass bolt.

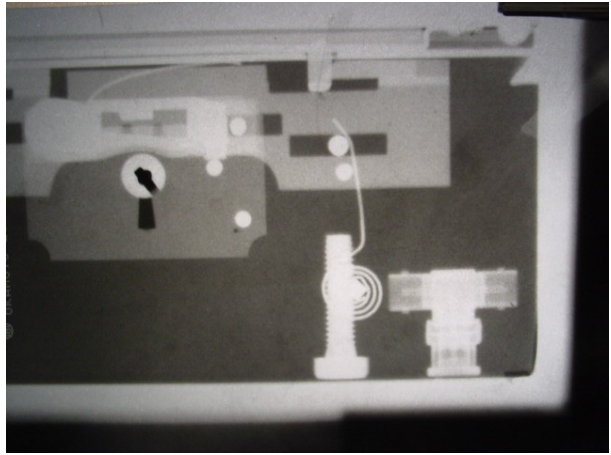

**Figure 14**
Radiography of the strongbox shown in figure 13.

## Acknowledgements
This research was supported by PLADEMA – CNEA, and Universidad de Buenos Aires. VR, FDL, PK and AT are doctoral fellows of CONICET. CM and AC are members of CONICET.